# Experimental search for the permanent electric dipole moment of an Na atom using special capacitor in the shape of Dewar flask

Pei-Lin You

Institute of Quantum Electronics, Guangdong Ocean University, zhanjiang 524025, China.

Since the time of Rutherford it was commonly believed that with no electric field, the nucleus of an atom is at the centre of the electron cloud, so that all kinds of atoms do not have permanent electric dipole moment(EDM). In the fact, the idea is untested hypothesis. Using two special capacitors containing Sodium vapor we find the electric susceptibility $x_e$ of Na atoms is directly proportional to its density N, and inversely to the absolute temperature T, as polar molecules. $x_e$=A+B/T, where A≈0, B=126.6±3.8(K) and N=1.49×$10^{20}$ m$^{-3}$. A ground state neutral Na atom has a large permanent EDM: $d_{Na}$=[1.28±0.15(stat)±0.10(syst)]×$10^{-8}$e.cm. The non-zero observation of EDM in any non-degenerate system will be a direct proof of time-reversal violation in nature, and new example of CP violation occurred in Na atoms. We work out the most linear Stark shift of Na atoms is only 0.0033nm, and so its linear Stark effect has not been observed till now! The experimental Na material with purity 0.9995 is supplied by Strem Chemicals Co. USA. The previous measurements have given $d_K$, $d_{Rb}$ and $d_{Cs}$. These results can be repeated in other laboratories, we welcome anyone who is interested in the experiments to visit and examine them.

**1. Introduction**   The realization that there is one small nucleus, which contains the entire positive charge and almost the entire mass of the atom, is due to the investigations of Rutherford, who utilized the scattering of alpha particles by matter. Since the time of Rutherford it was commonly believed that in the absence of an external field, the nucleus of an atom is at the centre of the electron cloud, so that any kind of atoms does not have permanent electric dipole moment (EDM)[1]. Therefore, there is no polar atom in nature except for polar molecules. In the fact, the idea is untested hypothesis. We hope thoroughly to test the idea by precise measurement. The experimental Sodium material with purity 99.95％ is supplied by Strem Chemicals Co.USA(see Fig.7). Our experiments showed that with no electric field, the centre of the electron cloud of Na atom does not coincide with its nucleus, and ground state neutral Na atom has a large EDM. This result is the product of past ten years of intense research [14-23].

In order for an atom or elementary particle to possess a EDM, time reversal (T) symmetry must be violated, and through the CPT theorem CP(charge conjugation and parity) must be violated as well[1]. The non-zero observation of EDM in any non-degenerate system will be a direct proof of time-reversal violation in nature[1-5]. On the other hand, some evidence for CP violation beyond the Standard Model comes from Cosmology. Astronomical observations indicate that our Universe is mostly made of matter and contains almost no anti-matter. The first example of CP violation was discovered in 1964, but it has been observed only in the decays of the $K_o$ mesons. After 38 years, the BaBar experiment at Stanford Linear Accelerator Center (SLAC) and the Belle collaboration at the KEK laboratory in Japan announced the second example of CP violation. "The results gave clear evidence for CP violation in B mesons. However, the degree of CP violation now confirmed is not enough on its own to account for the matter-antimatter imbalance in the Universe." (SLAC Press Release July 23, 2002). "In the past few years we have learned to cleanly isolate one event in a million." (SLAC Press Release Sep 28, 2006).Theorists found it hard to see why CP symmetry should be broken at all and even harder to understand why any imperfection should be so small[5]. This fact suggests that there must be other ways in which CP symmetry breaks down and more subtle effects must be examined. So EDM experiments are now considered an ideal probe for evidence of new sources of CP violation. If an EDM is found, it will be compelling evidence for the existence of new sources of CP violation.

Experimental searches for EDMs can be divided into three categories: search for the neutron EDM (the new result is $d_n$<2.9×$10^{-26}$e.cm)[2], search for the electron EDM utilizing paramagnetic atoms or molecule, the most sensitive of which is done with Tl atoms(the new result is $d_e$=[6.9±7.4]×$10^{-28}$e.cm)[3], and search for an EDM of diamagnetic atoms, the most sensitive of which is done with $^{199}$Hg(the new result is $d_{Hg}$=[0.49±1.29(stat)±0.76 (syst)]×$10^{-29}$e.cm )[1].

In fact, despite the relentless search for a non-zero EDM of atom or elementary particle for more 50 years, no conclusive results have been obtained so far[1-4]. In most experiments, they measured microcosmic Larmor precession frequency of individual particle based on nuclear spin or electron spin. The search for an EDM consists of measuring the precession frequency of the particle in parallel electric and magnetic fields and looking for a change of this frequency when the direction of **E** is reversed relative to **B.** We now submit the article on the similar topic, however, with measuring macroscopic electric susceptibility($x_e$) of Na vapor containing a large number of Na atoms (the density N >10$^{20}$m$^{-3}$). This article reported three new experimental phenomenon which have not been observed till now.

Our experiments showed that new example of CP violation occurred in Na atoms and it is a classic example of how understanding of our Universe advances through atomic physics research. These results can be repeated in other laboratories. Our experimental apparatus are still kept, we welcome anyone who is interested in the experiments to visit and examine it.

Madame Chien-Shiung Wu Completed the famous beta-decay of cobalt-60 experiment in 1957. It established violation of parity in weak interaction. Her experimental apparatus is just a glass Dewar flask containing a large number of cobalt-60 atoms. Everybody expected Madame Wu to find a left-right symmetry in the process of beta-decay. But nature sprang another surprise! The two tests were very much alike not only in the experimental apparatus and topic but also in surprising result.

R.P. Feynman considered a polar gas whose molecules have a permanent dipole moment $d_o$, such as water vapor [6]. With no electric field, these dipole moments are oriented at random, and no macroscopic dipole moment is observed. But when an electric field is applied, the electric dipoles tend to orient in the direction of the field. On the other hand, when atoms are placed in an electric field, they become polarized, acquiring induced electric dipole moments in the direction of the field. Note that the electric susceptibility($x_e$) caused by the orientation of polar molecules is inversely proportional to the absolute temperature(T): $x_e$= B/T while the induced electric susceptibility due to the distortion of electronic motion in atoms or molecules is temperature independent: $x_e$=A, where A and the slope B is constant. The electric susceptibility $x_e$ =C/C$_o$ - 1, C$_o$ is the vacuum capacitance and C is the capacitance of the capacitor filled with the material. This difference in temperature dependence offers a means of separating the polar and non-polar substances experimentally[6].

The electric susceptibility caused by the orientation of polar molecules is [6]

$$x_e = N d_o^2 / 3kT \varepsilon_o \qquad (1)$$

where k = 1.38066×10$^{-23}$ J/K is Boltzmann constant, $\varepsilon_o$ = 8.8542×10$^{-12}$ A/V. m is the permittivity of free space, N is the number density of molecules. R.P. Feynman checked Eq.(1) with the orientation polarization experiment of water vapor. He plotted the straight line from four experimental points. The table 1 gives the experimental data [6].

**Table 1    The electric susceptibility $x_e$ of Water vapor at different temperature T**

| t (℃) | T (K) | 1/T(K$^{-1}$) | $x_e$ | P(cmHg) | N(m$^{-3}$) |
|---|---|---|---|---|---|
| 120 | 393.15 | 2.5436×10$^{-3}$ | 400.2×10$^{-5}$ | 56.49 | 1.388×10$^{25}$ |
| 150 | 423.15 | 2.3632×10$^{-3}$ | 371.7×10$^{-5}$ | 60.93 | 1.391×10$^{25}$ |
| 180 | 453.15 | 2.2068×10$^{-3}$ | 348.8×10$^{-5}$ | 65.34 | 1.393×10$^{25}$ |
| 210 | 483.15 | 2.0698×10$^{-3}$ | 328.7×10$^{-5}$ | 69.75 | 1.395×10$^{25}$ |

From the ideal gas law, the average density N= P/ k T =1.392×10$^{25}$ m$^{-3}$. From least-square method we obtain A=1.8×10$^{-4}$≈0, B=1.50K, and the permanent EDM of a H$_2$O molecule $d_{H2O}$ = (3k$\varepsilon_o$B /N )$^{1/2}$ = 6.28×10$^{-30}$ C.m = 0.393×10$^{-8}$e. cm, it is conform to the observed value $d_{H2O}$= 6.20×10$^{-30}$ C. m=0.388×10$^{-8}$e. cm [7]**.** By measuring capacitance at different temperatures, it is possible to distinguish between permanent and induced dipole moments[7]. If Na atom is the polar atom and has a large EDM, a temperature dependence of the form $x_e$=B/T should be expected in measuring the capacitance.

**2. Experimental method and result**    The first experiment: investigation of the relationship between $x_e$ of Na vapor and its number density N. The experimental apparatus is a closed glass container. It resembles a Dewar flask in shape. The external and internal diameters of the container are D$_1$=78.5mm and D$_2$=54.5mm. The external and internal surfaces of the container are plated with silver, respectively shown by **a** and **b** in Fig.1. These two silver layers constitute the cylindrical capacitor. The length of the two silver layers is L=26.5cm. The thickness of the glass wall is △=1.5mm. The width of the gap that will be filled by Na vapor is H$_1$=9.0mm. The magnitude of capacitance is measured by a digital capacitance meter. The precision of the meter was 0.1pF, the accuracy was 0.5% and the surveying voltage was V=1.2 volt. This capacitor is connected in series by two capacitors. One is called C' and contains the Na vapor of thickness H$_1$, another one is called C'' and contains the glass medium of thickness 2△. The total capacitance C is

$$C = C'C''/(C'+C'') \quad \text{or} \quad C' = C''C/(C''-C) \qquad (2)$$

where C'' and C can be directly measured. The experiment to measure C'' is easy. We can made a cylindrical capacitor of glass with thickness 2△ and put it in a temperature-control stove. By measuring capacitances at different temperatures, we can find C'' correspond to different temperatures. When the container is empty, it is pumped to a vacuum pressure P≤$10^{-8}$ Pa for 20 hours. We measured the total capacitance C = 52.4pF and C'' = 1632pF, the vacuum capacitance is C'$_0$ =54.1 pF according to (2). The mass of a sodium sample, with purity 0.9995, is 5g and supply by Strem chemicals company USA(Fig.7). The next step, the Na sample is put in the container. The container is again pumped to vacuum pressure P ≤$10^{-8}$ Pa at room temperature, then it is sealed. We obtain the experimental apparatus as shown in Fig.1 and it is a glass Dewar flask filled with Na vapor and surplus liquid sample. We put the capacitor into the temperature-control stove, raise the temperature of the stove very slow and keep the temperature at $T_1$ =591.15K(318℃) for 4 hours. The electric susceptibility $x_e$ of Na vapor is measured at different time when the temperature keeps constant $T_1$. The experimental curve is shown in Fig.2

The experiment showed that with prolongation of isothermal time and volatilization of surplus liquid sodium, both N and $x_e$ of Na vapor simultaneously increases rapidly. When t≥180 minutes at $T_1$ =591.15K, both the total capacitance and glass capacitance remain constants: $C_t$= 1468 pF and C'' = 8160 pF. And the capacitance of Na vapor is C'$_t$=1790 pF according to (2). It means that the readings of capacitance are obtained under the condition of Na saturated vapor pressure. So the electric susceptibility of Na vapor remains constant $x_e$ = C'$_t$/ C'$_0$ - 1=32.1. The formula of saturated vapor pressure of Na atoms is P= $10^{7.553-5395.4/T}$ psi, where 1 psi =6894.8Pa[8]. The effective range of the formula is 453K ≤T ≤1156K. Using the formula we obtain the saturated vapor pressure of Na vapor $P_{Na}$=183.9Pa at 591.15K. From the ideal gas law, the density of Na vapor $N_1$ = $P_{Na}$ /k$T_1$ =2.25×$10^{22}$ m$^{-3}$. Note that when the isothermal time t≤60 minutes, the plot is a straight line. Because the density N of Na vapor is proportional to the volatile time t, the straight line shows that the electric susceptibility $x_e$ of Na vapor is directly proportional to its density N when T keeps constant.

It is well known that the electric susceptibility is of the order of $10^{-3}$ for any kind of gases, for example 0.0046 for HCl gas, 0.007 for water vapor [7]. Please notice that $x_e$ = 32.1>>1 for Na vapor and the digital meter applied the external field only with E=V/ $H_1$=1.33×$10^2$V/m, and it is very weak. Our experimental result exceeded all expert's expectation!

The second experiment: investigation of the relationship between $x_e$ of Na vapor and T at a fixed density. The apparatus was a closed glass container but the Na vapor was at a fixed density $N_2$. In order to control the quantity of Na vapor, the container is connected to another small container that contains Na material by a glass tube from the top. These two containers are slowly heated to $T_2$=473.15K(200℃) in the stove for 3 hours and the designed experimental container is sealed. The apparatus shown in Fig.3 and two stainless steel tubes a and b build up the cylindrical capacitor. The external and internal diameters of the two tubes are $D_3$=67.4mm and $D_4$=50.4mm and its length is $L_2$=23.0 cm. The plate separation is $H_2$=8.5mm. The capacitance C was still measured by the digital meter and the vacuum capacitance $C_{20}$=44.0pF. Using the formula of saturated vapor pressure of Na atoms we obtain $P_{Na}$=0.9736 Pa at $T_2$=473.15K. From the ideal gas law, the density of Na atoms $N_2$= $P_{Na}$ /k$T_2$ =1.49×$10^{20}$ m$^{-3}$. By measuring electric susceptibility of Na vapor at different temperature, we obtain $x_e$ =A+B/T≈B/T, where the intercept A≈0 and the slope of the line B=126.6±3.8 (K). The experimental results are shown in Fig.4. Table 2 gives a complete experimental data.

**Table 2 The electric susceptibility $x_e$ of Na vapor at different temperature T**

| t (℃) | 114 | 122 | 130 | 138 | 146 | 154 | 164 | 174 | 186 |
|---|---|---|---|---|---|---|---|---|---|
| T(K) | 387.15 | 395.15 | 403.15 | 411.15 | 419.15 | 427.15 | 437.15 | 447.15 | 459.15 |
| 1/T(×$10^{-3}$) | 2.5830 | 2.5307 | 2.4805 | 2.4322 | 2.3858 | 2.3411 | 2.2875 | 2.2364 | 2.1779 |
| C(pF) | 58.8 | 58.5 | 58.3 | 58.0 | 57.7 | 57.4 | 57.1 | 56.9 | 56.6 |
| $x_e$ | 0.33636 | 0.32955 | 0.3250 | 0.31818 | 0.31136 | 0.30455 | 0.29773 | 0.29318 | 0.28636 |

where $C_{20}$= 44.0pF, $N_2$=1.49×$10^{20}$ m$^{-3}$. From least-square method we obtain B= 126.6K and A=0.0096≈0, where the surveying voltage V=1.2 volt. The digital meter applied the external field only with E=V/ $H_2$=1.4×$10^2$V/m and it is very weak. From Eq. (11) we work out $d_{Na}$ = 2.048×$10^{-29}$C.m= 1.280×$10^{-8}$e.cm＞$d_{H2O}$!

The third experiment: measuring the capacitance of Na vapor at various voltages (V) under the fixed density $N_2$ and a fixed temperature $T_3$. The apparatus was the same as the preceding one and $C_{30}= C_{20}$=44.0 pF. The measuring method is shown in Fig. 5. C was the capacitor filled with Na vapor to be measured and kept at $T_3$ =303.15K (30℃). $C_d$ =520pF was used as a standard capacitor. Two signals $V_c(t)=V_{co} \cos \omega t$ and $V_s(t)=V_{so} \cos \omega t$ were measured by a two channel digital real-time oscilloscope (supply by Tektronix TDS 210 USA). The two signals had the same frequency and always the same phase at different voltages. From Fig.5, we have $(V_s-V_c)/V_c=C/C_d$ and $C=(V_{so}/V_{co} - 1)C_d$. In the experiment $V_{so}$ can be adjusted from zero to 800V. When $V_1=V_{co}$≤0.4volt, $C_1$=216pF or $x_e$ =3.91, it is constant. With the increase of voltage, the capacitance decreases gradually. When $V_2=V_{co}$=400 volt, $C_2$=47.0pF or $x_e$= 0.0682, it approaches saturation. The measured result of the electric susceptibility at different voltages is shown in Fig.6. The $x_e$-V curve showed that the saturation polarization of the Na vapor is obvious when $E$≥$V_2/ H_2$=4.7×$10^4$V/m(see discuss ⑥).

**3. Theory and interpretation**    The local field acting on a molecule in a gas is almost the same as the external field **E**[7]. The electric susceptibility of a gaseous polar dielectric is[9]

$$x_e = NG+ N d_o L(a)/ \varepsilon_o E \tag{3}$$

where $a = d_o E /kT$, $d_o$ is EDM of a molecule, G is the molecule polarizability. The Langevin function is

$$L(a) = [(e^a + e^{-a}) / (e^a – e^{-a})] – 1/a \tag{4}$$

The Langevin function L(a) is equal to the mean value of $\cos \theta$ ( $\theta$ is the angle between $\mathbf{d_o}$ and **E**) [9]:

$$<\cos \theta> = \mu \int_0^\pi \cos \theta \exp (d_o E \cos \theta /kT) \sin \theta \, d\theta= L(a), \quad \mu =[\int_0^\pi \exp (d_o E \cos \theta /kT) \sin \theta \, d\theta]^{-1} \tag{5}$$

where $\mu$ is a normalized constant. This result shows that L(a) is the percentage of polar molecule lined up with the field in the total number. When a<<1 and L(a)≈a/3, when a>>1 and L(a)≈1[9].

The next step, we will consider how this equation is applied to Na atoms. Due to the atomic polarizability of Na atoms is G= 24.1×$10^{-30}$ $m^3$[10], the number density of Na atoms N≤$10^{23}$ $m^{-3}$ and the induced susceptibility A=NG≤2.41×$10^{-6}$ can be neglected. In addition, the induced dipole moment of Na atoms is $d_{int}$ =G $\varepsilon_o$ E [10], due to E≤$10^5$v/m in the experiment, then $d_{int}$≤2.13×$10^{-35}$ C.m can be neglected. From Eq.(3) we obtain

$$x_e = Nd L(a)/ \varepsilon_o E \tag{6}$$

where d is the EDM of an Na atom and N is the number density of Na vapor. L(a)= $<\cos \theta>$ is the percentage of Na atoms lined up with the field in the total number. Note that E=V/H and $\varepsilon_o= C_o H / S$, leading to

$$C - C_o = \beta L(a)/a \quad \text{or} \quad x_e = \beta L(a)/a \, C_o \tag{7}$$

where $\beta = S N d^2/kTH$ is a constant. **This is the polarization equation of Na atoms**. Due to a=d E/kT= dV/kTH we obtain **the first formula of atomic EDM**

$$d_{atom} =(C - C_o )V / L(a)SN \tag{8}$$

In order to work out **L(a)** and **a** of the first experiment, note that in the third experiment when the field is weak ($V_1$=0.4V), $a_1$ <<1 and $L(a_1)$≈$a_1$/3. From Eq.(7): $C_1 - C_{30}$=216 - 44= $\beta$/3 and $\beta$ =516pF. When the field is strong ($V_2$=400V), $a_2$ >>1 and $L(a_2)$≈1, $C_2 - C_{30} = L(a_2) \beta /a_2$. We work out $a_2 = \beta L(a_2)/(C_2 - C_{30})$= 171>>1, $L(a_2)$=0.9942. Due to a=dV/kTH, so $a/a_2$=$VT_2H_2/T_1H_1V_2$, Substituting the corresponding values, **a**= 0.2485 and L(a)=0.082. L(a)=0.082 shows that only 8.2％ of Na atoms are lined up with the direction of the field in the first experiment. Notice that we deduced Eq(7) and Eq(8) from the formula of the parallel-plate capacitor $\varepsilon_o= C_o H / S$, so the cylindrical capacitor must be regarded as a equivalent parallel-plate capacitor with the plate area S= $C_o H/ \varepsilon_o$ In the first experiment the equivalent plate area $S_1$= $C'_0 H_1/ \varepsilon_o$ =5.5×$10^{-2}$ $m^2$. Substituting the values: $S_1$= 5.5×$10^{-2}$ $m^2$, $N_1$ =2.25×$10^{22}$ $m^{-3}$, V=1.2volt and $C - C_o = C'_t - C'_0$ = 1735.9pF, from Eq. (8) we work out

$$d_{Na} =(C - C_o )V / L(a) S_1 N_1 =2.053 \times 10^{-29} C.m = 1.283 \times 10^{-8} e.cm \tag{9}$$

The statistical error about the measured value is △$d_1$/d=△C/C+△$C_o$/$C_o$+△$S_1$/$S_1$+△V/V+△$N_1$/$N_1$＜0.12, considering all sources of systematic error is △$d_2$/d＜0.08, and the combination error is △d/d＜0.15. We find that

$$d_{Na}=[2.05 \pm 0.25(\text{stat}) \pm 0.16 (\text{syst})] \times 10^{-29} C.m = [1.28 \pm 0.15(\text{stat}) \pm 0.10 (\text{syst})] \times 10^{-8} e. cm \tag{10}$$

Although above calculation is simple, but no physicist completed the calculation up to now!

From d= $(3k \varepsilon_o B / N_2 )^{1/2}$, note that k= dE/a$T_1$= dV/a$T_1 H_1$, we obtain **the second formula of atomic EDM**

$$d_{atom}= (3k \varepsilon_o B / N_2)^{1/2}= 3 V \varepsilon_o B/ aT_1 H_1 N_2 \tag{11}$$

Substituting the values: $T_1$= 591.15k, $H_1$ =9mm, V=1.2volt, B=126.6K, a=0.2485 and $N_2$ =1.490×$10^{20}$ $m^{-3}$, from

Eq. (11) we work out

$$d_{Na} = 3V\varepsilon_o B / aT_1 H_1 N_2 = 2.048 \times 10^{-29} C.m = 1.280 \times 10^{-8} e.cm \qquad (12)$$

**Using two different methods and different experimental data we obtain the same result, it proved that the data are reliable and the EDM of an Na atom has been measured accurately.**

## 4. Discussion

①The formula $d_{atom} = (C - C_o)V/L(a)SN$ can be justified easily. The dipole moment of an Na atom is $d = er$. N is the number of Na atoms per unit volume. L(a) is the percentage of Na atoms lined up with the field in the total number. When an electric field is applied, the Na atoms tend to orient in the direction of the field as dipoles. On the one hand, the change of the charge of the capacitor is $\triangle Q = (C - C_0)V$. On the other hand, due to the volume of the capacitor is SH, the total number of Na atoms lined up with the field is SHNL(a). The number of layers of Na atoms which lined up with the filed is H/r. Because inside the Na vapor the positive and negative charges cancel out each other, the polarization only gives rise to a net positive charge on one side of the capacitor and a net negative charge on the opposite side. Then the change of the charge is $\triangle Q = SHNL(a)e / (H/r) = SNL(a)d = (C - C_0)V$, so the EDM of an Na atom is $d = (C - C_0)V / SNL(a)$.

②If Na atom has a large EDM, why the linear Stark effect has not been observed? This is an interesting question. As a concrete example, let us treat the fine structure and the linear Stark shifts of the hydrogen( n=2). Notice that the fine structure of the hydrogen (n=2) is only 0.33 cm$^{-1}$ for the Hα lines of the Balmer series, where $\lambda = 656.3$ nm, and the splitting is only $\triangle \lambda = 0.33 \times (656.3 \times 10^{-7})^2 = 0.014$ nm, therefore the fine structure is difficult to observe [11]. The linear Stark shifts of the energy levels is proportional to the field strength: $\triangle W = d_H E = 1.59 \times 10^{-8} E$ e.cm. When $E = 10^5$ V/cm, $\triangle W = 1.59 \times 10^{-3}$ eV, this corresponds to a wavenumber of 12.8 cm$^{-1}$. So the linear Stark shifts is $\triangle \lambda = \triangle W \lambda^2 / hc = 12.8 \times (656.3 \times 10^{-7})^2 = 0.55$ nm. It is so large, in fact, that the Stark shift of the lines of the hydrogen is easily observed [11]. However, the most field strength for Na atoms is $E_{max} = 4.7 \times 10^4$ V/m, if Na atom has a large EDM $d_{Na} = 1.28 \times 10^{-8}$ e.cm, and the most splitting of the energy levels of Na atoms $\triangle W_{max} = d_{Na} E_{max} = 6.02 \times 10^{-6}$ eV. This corresponds to a wavenumber of $4.85 \times 10^{-2}$ cm$^{-1}$. On the other hand, the observed values for a line pair of the first primary series of Na atom( Z=11, n=3) are $\lambda_1 = 819.48$ nm and $\lambda_2 = 818.33$ nm[8]. The magnitude of the linear Stark shift of Na atoms is only $\triangle \lambda = \triangle W (\lambda_1 + \lambda_2)^2 / 4hc = 0.0033$ nm. **It is so small, in fact, that a direct observation of the linear Stark shifts of Na atom is not possible!**

③The shift in the energy levels of an atom in an electric field is known as the Stark effect. Normally the effect is quadratic in the field strength, but first excited state of the hydrogen atom exhibits an effect that is linear in the strength. This is due to the degeneracy of the excited state. This result shows that the hydrogen atom (the quantum number n=2 ) has very large EDM, $d_H = 3e\,a_o = 1.59 \times 10^{-8}$ e.cm ($a_o$ is Bohr radius) [12]. L.D. Landay once stated that "The presence of the linear effect means that, in the unperturbed state, the hydrogen atom has a dipole moment" [12]. The alkali atoms having only one valence electron in the outermost shell can be described as hydrogen-like atoms[13]. Since the quantum number of the ground state alkali atoms are n≥2 rather than n=1( this is 2 for Li, 3 for Na, 4 for K, 5 for Rb and 6 for Cs), as the excited state of the hydrogen atom. So we conjecture that the ground state neutral alkali atoms may have large EDM of the order of $e\,a_o$ [14]. Due to the EDM of an atom is extremely small and we applied several ingenious experimental techniques [15].

④From 1ev=kT we get T=11594K, in the temperature range of the experiment 303K ≤T≤591K, kT<<1ev, so the measured capacitance change ( $C'_t - C'_0$ ) entirely comes from the contribution of ground state Na atoms.

⑤If nearly all the dipoles in a gas turn toward the direction of the field, then θ = 0, and the mean value of cos θ <cos θ> = L(a)=1, this effect is called the saturation polarization. Because the most EDM $d_{omax}$ of polar molecules is roughly $10^{-29}$ C.m and the breakdown field intensity of gaseous dielectric is roughly $10^7$ v/m. The most potential energy $d_{omax}E$ of a molecule is about $6.3 \times 10^{-4}$ ev. The average kinetic energy kT of each molecule in a gas is about 0.03ev at ordinary temperatures(T=300k), and most value $a_{max} = d_{omax}E/kT = 0.021 <<1$ despite under a breakdown field. So human have never observed the saturation polarization effect of any gaseous dielectric at ordinary temperatures till now. R.P. Feynman once stated that *" when a filed is applied, if all the dipoles in a gas were to line up, there would be a very large polarization, but that does not happen"* [6].

⑥The polarization P, the dipole moment per unit volume, is

$$P = \varepsilon_o \chi_e E \qquad (13)$$

When $V_2 = 400$ volt, $a_2 = 171$, from Eq.(7) $\chi_e = L(a_2)\beta / a_2 C_{20}$, substituting $\chi_e$, $\varepsilon_o$ and $E_2$ in Eq.(13) we obtain

$$P_2 = \varepsilon_o \chi_e E_2 = (C_2 - C_{20})V_2 / S = Nd\,L(a_2) = 0.9942\,Nd \approx N\,d \qquad (14)$$

We obtain a very large polarization! When $V_1=0.4$ volt, $a_1= a_2V_1/V_2 =0.171$, $L(a_1)= a_1/3=0.057$ and $\chi_e= \beta /3C_{20}$, substituting $\chi_e$, $\varepsilon_o$ and $E_1$ in Eq. (13) we obtain

$$P_1 = \varepsilon_o \chi_e E_1=(C_1 - C_{20})V_1/ S = Nd\, L(a_1)= Nd\, a_1/3=0.057\,Nd \approx Nd /18 \qquad (15)$$

Evidently this polarization is very small. $L(a_2)=0.9942$ shows that nearly all Na atoms (more than 99.4％) are lined up with the direction of the field when $V_2 \geq 400$ v. It showed that the saturation polarization of the Na vapor is obvious when $E \geq 4.7\times 10^4$V/m It tells us that the saturation polarization of Na vapor at ordinary temperatures is an entirely unexpected discovery and another two articles discussed the question [21, 22].

⑦Accurate measurements of the EDM of an Cesium (Cs), Rubidium (Rb) and Potassium(K) atom in ground state have been carried out. Similar results have been obtained as follows.

$d_K = [1.58 \pm 0.19 \text{(stat)} \pm 0.13 \text{(syst)}]\times 10^{-8}$ e. cm [23]

$d_{Rb} = [1.70 \pm 0.20 \text{(stat)} \pm 0.14 \text{(syst)}]\times 10^{-8}$ e. cm [20]

$d_{Cs} = [1.86 \pm 0.22 \text{(stat)} \pm 0.15 \text{(syst)}]\times 10^{-8}$ e. cm [19]

In conclusion, we demonstrate a new protocol to search for the permanent electric dipole moment of single alkali atom such as Na, K. Rb and Cs atom. Clearly, realization of this protocol poses considerable experimental challenges and opens up exciting opportunities to find new sources of CP violation. We felt a shock even greater than Cronin, Fitch, et. al. have felt when they first encountered CP violation in the decays of the $K_o$ mesons in 1964. We completed the test which our more symmetric minded Western physical colleagues had thought scarcely worth the effort! The history of science can be described as a continual and never-ending discovery of new problems. Perhaps, it is the most important result in atomic physics since the experiment of Rutherford and his colleagues in 1909. Few experiments have produced a result as surprising as this one.

**Acnowledgement**

The author thank to Prof. Wei-Min Du, Dr. Yo-Sheng Zhang, thank to Engineer Yi-Quan Zhan, Engineer Jia You for their help in the work.

**References**

1. W C. Griffith, M.D.Swallows, T.H.Loftus, M. V. Romalis, B.R.Heckel, and E N. Fortson, Phys. Rev. Lett. **102,** 101601 (2009)
2. C. A. Baker *et al*., Phys. Rev. Lett. **97,** 131801 (2006)
3. B.C. Regan *et al*., Phys .Rev. Lett. **88,** 071805 (2002)
4. M. V. Romalis, W C. Griffith, J. P. Jacobs, and E. N. Fortson, Phys. Rev. Lett. **86,** 2505 (2001)
5. H. R. Quinn, and M. S. Witherell, Sci. Am., Oct. 1998, PP76-81
6. R.P.Feynman, R.B.Leighton, & M.Sands, The Feynman lectures on Physics Vol.2 (Addison-Wesley Publishing Co. 1964) P11.1-P11.5.
7. I.S.Grant, & W.R.Phillips, Electromagnetism. (John Wiley & Sons Ltd. 1975). PP57-65
8. J.A.Dean, Lange's Handbook of Chemistry (New York: McGraw-Hill ,Inc 1998)Table 5.3, Table 5.8
9. C.J.F.Bottcher, Theory of Electric Polarization (Amsterdam: Elsevier 1973) P161
10. D.R.Lide, Handbook of Chemistry and Physics. (Boca Raton New York: CRC Press, 1998) 10.201-10.202
11. H.Haken, and H.C.Wolf, The Physics of Atoms and Quanta. (Springer-Verlag Berlin Heidelberg 2000),P195,PP171-259
12. L.D.Landay, and E.M.Lifshitz, Quantum Mechanics(Non-relativistic Theory) (Beijing World Publishing Corporation 1999)P290,P2
13. W. Greiner, Quantum Mechanics an introduction(Springer-verlag Berlin/Heidelberg 1994) P213.
14. Pei-Lin You, Study on occurrence of induced and permanent electric dipole moment of a single atom, J. Zhanjiang Ocean Univ. Vol.**20** No.4, 60 (2000) (in Chinese)
15. Pei-Lin You, and Xiang-You Huang, Measurement of small capacitance with loss at low frequency, J. Data. Acquisition and Processing, Vol. **16** No. S, 30 (2001) (in Chinese)
16. Xiang-You Huang, and Pei-Lin You, Chin. Phys. Lett., **19** 1038 (2002)
17. Xiang-You Huang, and Pei-Lin You, Chin. Phys. Lett., **20** 1188 (2003)
18. Xiang-You Huang, Pei-Lin You, and Wei-Min Du, Chin. Phys., **13** 11 (2004)
19. Pei-Lin You, and Xiang-You Huang, arXiv: 0809.4767
20. Pei-Lin You, and Xiang-You Huang, arXiv: 0810.0770
21. Pei-Lin You, arXiv: 0812.3217
22. Pei-Lin You, arXiv: 0812.4424
23. Pei-Lin You, arXiv: 0908.3955(V2)

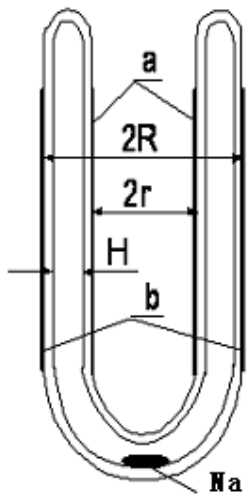

**Fig.1** The longitudinal section of the apparatus in the first experiment. The apparatus is a glass Dewar flask filled with Na vapor and surplus liquid sample. Silver layers **a** and **b** build up the cylindrical glass capacitor.

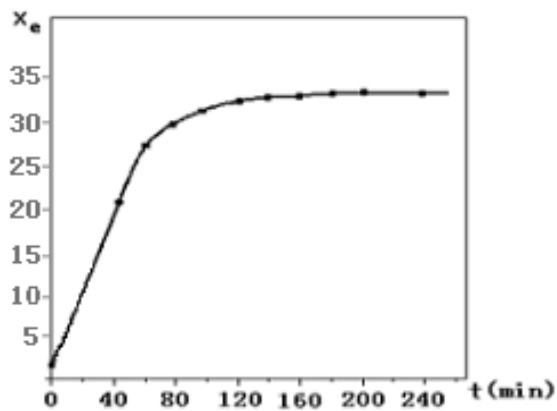

**Fig.2** The electric susceptibility $x_e$ of Na vapor is measured at different time when keeps at $T_1$=591K. Notice that $x_e = 32.1 \gg 1$ when $t \geq 180$ minutes, where the external field $E = 1.33 \times 10^2$ V/m, the density keeps at $N_1 = 2.25 \times 10^{22}$ m$^{-3}$.

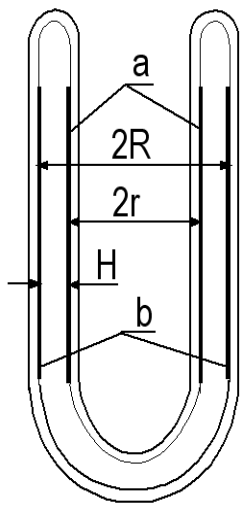

**Fig.3** The longitudinal section of the apparatus in another two experiments. The glass Dewar flask filled with Na vapor at a fixed density. Two stainless steel tubes **a** and **b** build up the cylindrical capacitor.

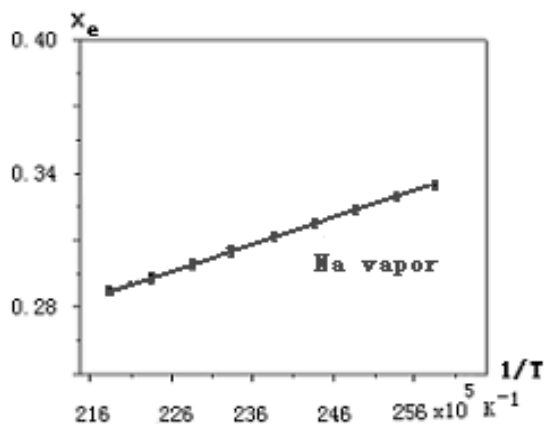

**Fig.4** The curve showed that $x_e$ of Na vapor varies inversely proportional to the absolute temperature T under a fixed density $N_2$. $x_e = A + B/T$, where A=0.0096, B=126.6(K) and the density $N_2$=1.49×10$^{20}$ m$^{-3}$.

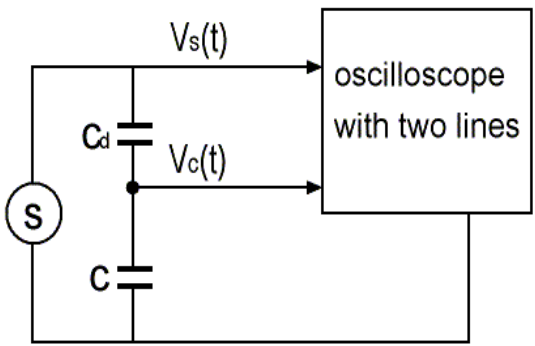
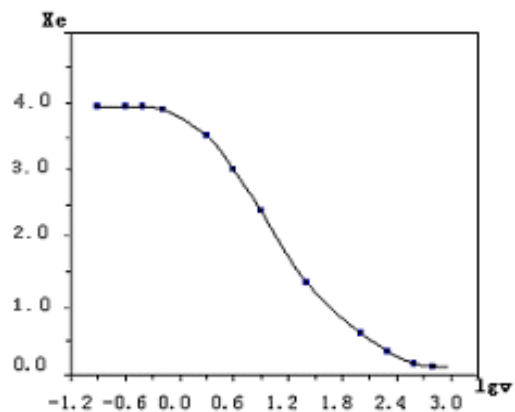

**Fig.5** The diagram shows the measuring method. C is a capacitor filled with Na vapor to be measured. Cd=520pF is used as a standard capacitor, where $V_s(t) = V_{so}\cos\omega t$ and $V_c(t) = V_{co}\cos\omega t$.

**Fig.6** The curve shows that the saturation polarization of the Na vapor is obvious when E $\geq 4.7\times 10^4$ v/m, where keeps at $T_3$ =303K, the density keeps at $N_2$=1.49×$10^{20}$ m$^{-3}$.

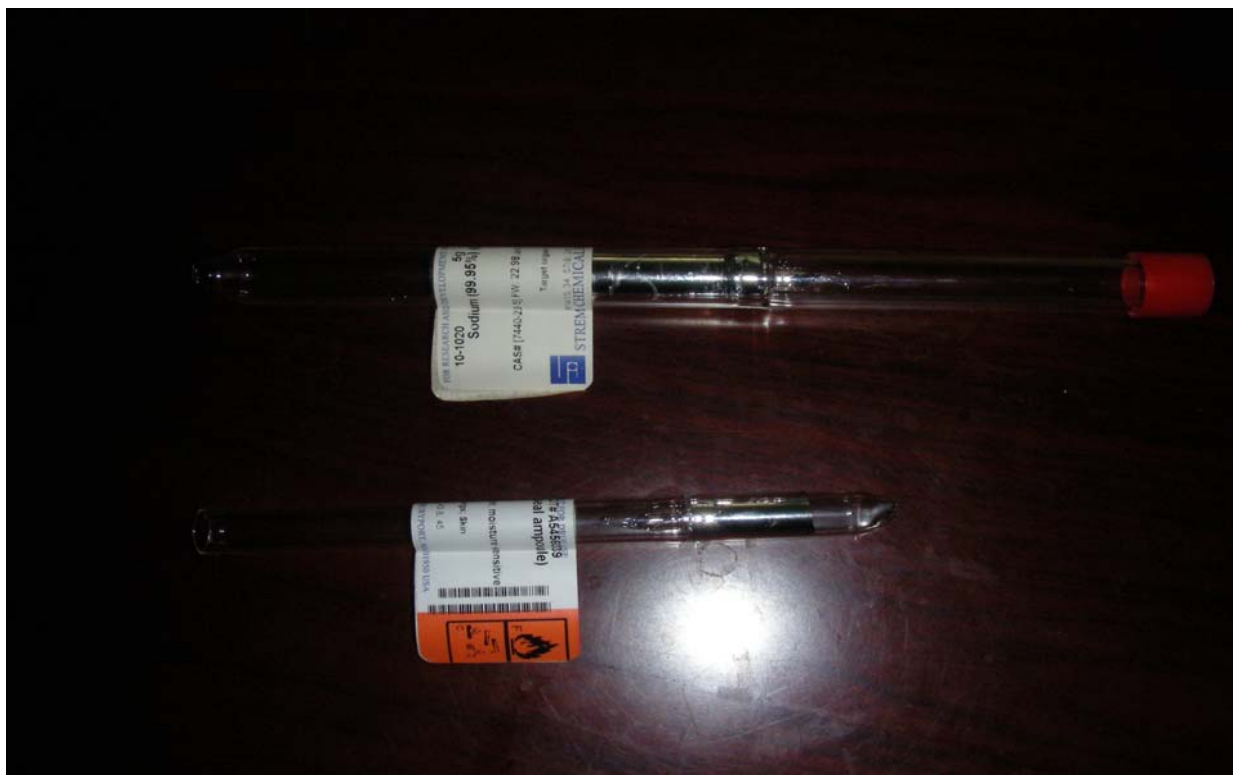

**Fig.7** The experimental Sodium material with purity 99.95％ is supplied by STREM CHEMICALS Co. USA. The mass of the two Sodium sample are 1g and 5g respectively. They are in their respective breakable ampoules.